# MULTIPACTORING CODE FOR 3D ACCELERATING STRUCTURES

L.V.Kravchuk, G.V.Romanov, S.G.Tarasov

Institute for Nuclear Research of RAS, 117312, Moscow Russia

*Abstract*

The simulation code has been developed to investigate possible electron multipacting in the RF cavities using preliminary calculated field components. The code provides different options of numerical study of the electron multipacting including search of the resonant trajectories and simulation of an electron multiplication.

## 1 INTRODUCTION

The new simulation code MultP has been developed on the base of the previous one [1] to investigate an electron multipacting in RF cavities using numerically calculated field components. The principal advantage of the MultP code is development of multipacting simulation procedure for 3D shape of RF cavity.

An easy and convenient interface along with a broad range of possibilities is considered as an important goal. As a whole the MultP is performed as a Windows application.

The MultP provides different options of numerical study of the electron multipacting. Particularly it is possible:
- to track separate electrons;
- to track a group of electrons;
- to scan electric field levels and initial phases of electron and to determine a possibility of resonance electron multipacting;
- to simulate multiplication of a number of randomly distributed electrons.

## 2 GENERAL DESCRIPTION

The program MultP uses 3D field map calculated by MAFIA. The files containing the field components are the files in standard text output format of MAFIA. The components may be calculated by other code and presented in mentioned format as well. These files are not used directly. MultP creates its own file containing interpolations of all field solutions to derive values E and B at the position of particles. This file is created once for given problem at first call of the program. During this procedure the distortions of field components near the curved surfaces, sharp edges, corners etc given by MAFIA are smoothed. Using the already pre-processed field data for subsequent runs of the program increases speed of calculations. The geometry of the investigated cavity is described by the user-written parameter-listing file.

Charged particle trajectory calculations in MultP are straightforward and involve numerical integration up to tens thousands of particles (limited by a computer power). Independent trajectory calculations are performed for each particle, so the code is applicable to the early stage of multipacting before space-charge effects become important or to the trajectories of separate particles. At each integration time step the MultP checks the status of the particles: are they inner or outer in respect to the cavity walls? The inner particle remains unimpeded. The outer particle is restored to its position just before entering the wall and is assigned a momentum corresponding to that of secondary electrons and directed oppositely. If the particles can not eject secondary electron or if the collision occurs during the decelerating phase of electromagnetic field the simulation for this particular particle terminates. A number of the secondary particles is random and depends on the secondary emission coefficient. The value of initial velocity of the secondary particle is also random inside interval selected by user.

## 3 DESCRIPTION OF THE OPTIONS

### 3.1 Tracking of the electrons

For the code options demonstration purposes the 350 MHz spoke cavity is used [2]. In the code the cavity is displayed as the views in different planes of Cartesian co-ordinate system (see Fig.1). The scaling in the views is distorted to fit the windows, but the windows can be opened outside of the main window with right ratio of the axes.

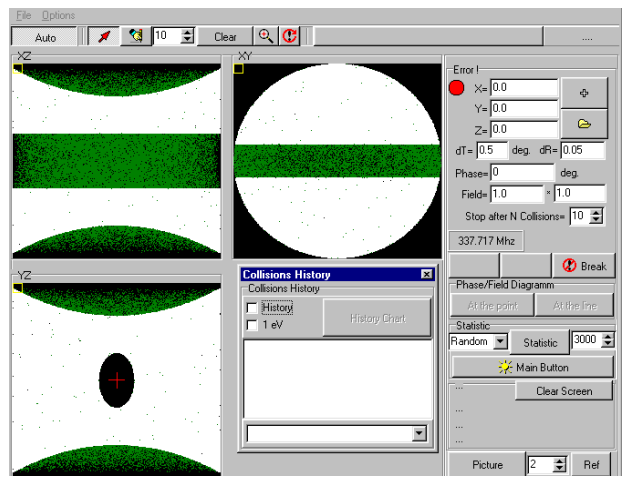

Fig.1. The main window of the code.

Though the code gives an opportunity to investigate a probability of the multipacting in the most general form (simulation of multiplication of randomly distributed electrons), but this way is rather time consuming. The tracking of separate electrons and group of electrons helps to grope the areas and the levels of RF field, which are favourable for multipacting.

To define the initial position of electron one can type its co-ordinates in the boxes or click mouse button at the positions in the views where one wants to put the particle. By default the program finds the nearest point on the surface and positions the particle at this point.

To observe a general character of electron motion in the area of interest it is convenient to launch a group of electrons with different initial positions. One has to define a number of particles and two points on a cavity wall. The line will be filled with particles.

After the initial velocities of electrons, the initial phase and the amplitude of the field are chosen in control boxes the trajectories are being calculated. The calculation of a trajectory terminates if number of collisions exceeds the predefined value or in a case of a negative phase.

The collision history window contains the information about the trajectories after each collision (Fig.2).

range from 1 to a value, specified by the user. The result is displayed as a phase/field diagram [3] shown in Fig.3. The red regions represent the trajectories that are the closest to the resonance. To exclude the trajectories with the energies outside the user defined range one should enable energy filter control (Fig.4).

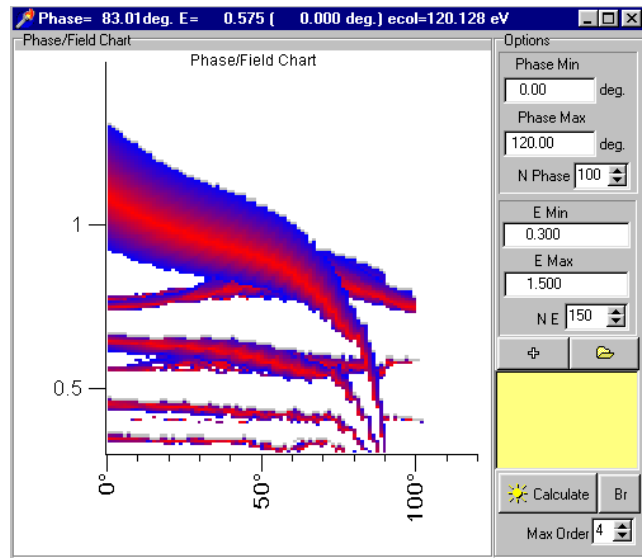

Fig.3. Phase/field diagram.

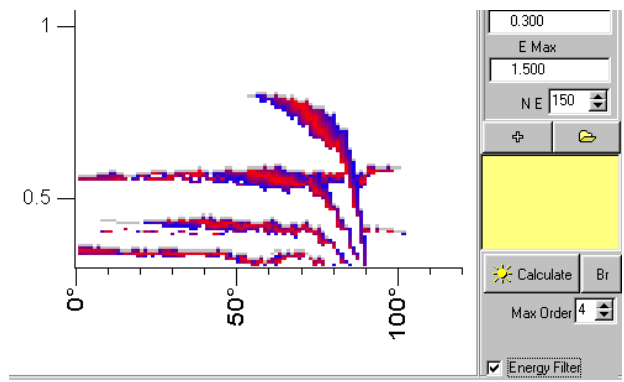

Fig.4. Phase/field diagram. Energy filter enabled.

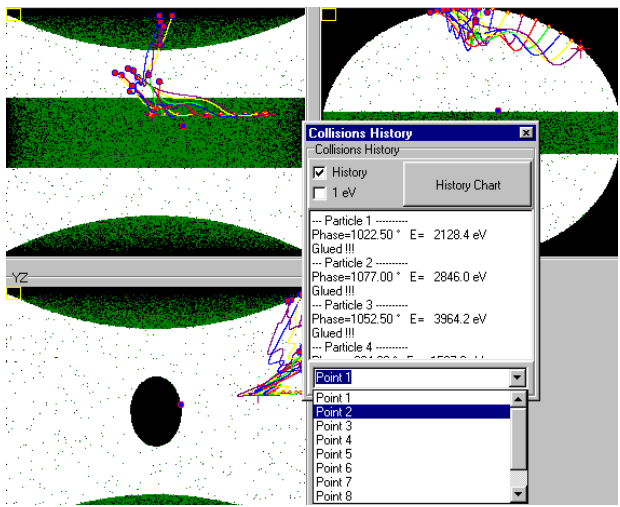

Fig.2. Trajectories of a group of electrons.

### 3.2 Phase/Field diagram

Tracking of a single electron or a group of electrons can help to find suspicious points inside a cavity. The phase/field diagram gives a certain answer to the question whether the resonant multiplication can exist at the suspected point.

For the point of interest one should define an interval of electron final energies in which the secondary-emission coefficient is greater than 1, the desired intervals and steps of scanning of phase and field levels. The program calculates a lot of trajectories and chooses among them the resonant ones for orders of multipacting within the

### 3.3 Direct simulation of multipacting

Straightforward simulation of multipacting is the most reliable way to check the existence of the multiplication process.

First of all the maximum value of the secondary emission coefficient and the interval of collision energies where this coefficient is greater than 1 are defined. During the simulation the secondary particles are launched if the incident electrons have an appropriate energy. The number of the secondary electrons is defined by the secondary emission coefficient. The initial velocities of the secondary electrons are randomly distributed in the interval from zero to user defined value.

The starting electrons (number is defined by user) are randomly distributed over all cavity or over chosen area. Their initial energies are randomly distributed in the interval defined by user and the initial phases are randomly distributed in the interval 0-120 degrees (this interval is fixed). The process is controlled with the use of windows showing the current number of particles vs number of rf cycles, phase distribution at the moment of collision, energy distribution at the moments of collision (see Fig.5, Fig.6 and Fig.7).

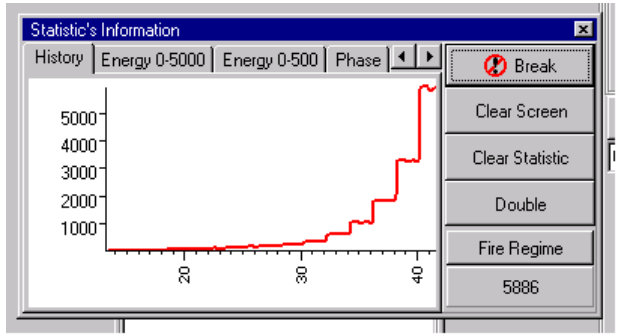

Fig.5. Multiplication of the electrons in the resonant discharge. Number of particles versus number of the cycles.

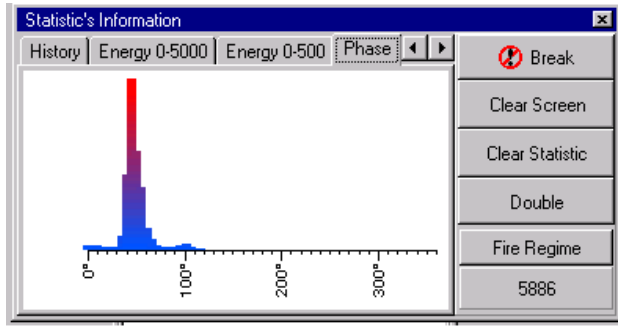

Fig.6. Equilibrium phase distribution at the moment of emission in the resonant discharge.

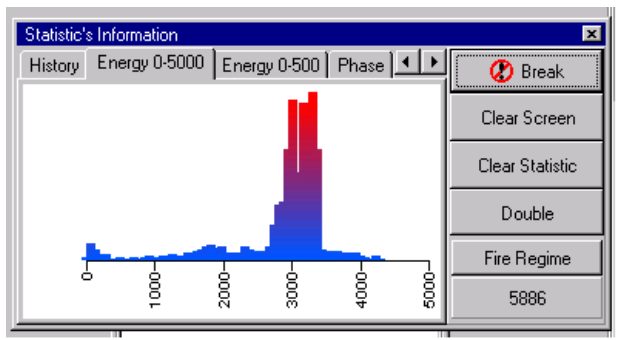

Fig.7. Equilibrium energy distribution of electrons at the moments of collision in resonant discharge.

Along with these control windows the program displays the current position of the particles as the moving points or as trajectories lines (see Fig. 8).

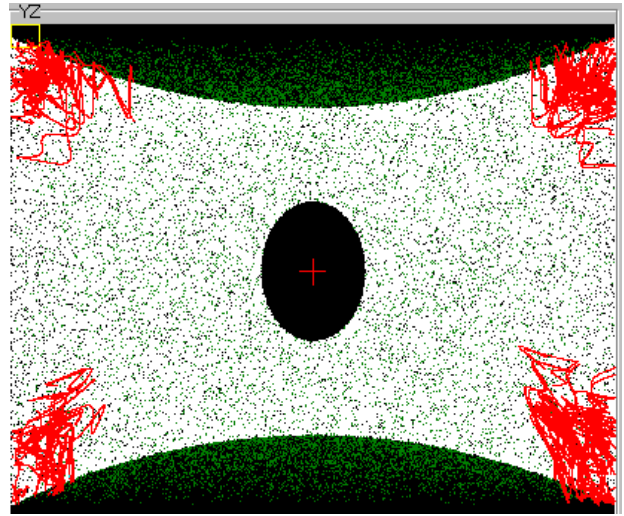

Fig.8. The electron trajectories in the non-resonant multipaction.

## CONCLUSION

The program "MultP" has been developed to investigate the electron multipacting in RF cavities using preliminary calculated field components. The main task was development of multipacting simulation procedure for realistic 3D shape of RF cavities. An easy and convenient interface along with a broad range of options were considered as an important goal.

The code uses MAFIA text files to generate 3D RF field map. A static magnetic field of uniform distribution can be applied in addition to RF fields.

So far the code uses the special user-written file to describe a cavity geometry. The geometry generator which will use MAFIA description of a cavity shape is being developed now.

The authors wish to acknowledge Alexander Feschenko for fruitful discussions and support.